\documentstyle[aps,preprint,tighten]{revtex}
\newcommand{\lsim}{\mathrel{\mathop{\kern 0pt \rlap
  {\raise.2ex\hbox{$<$}}}
  \lower.9ex\hbox{\kern-.190em $\sim$}}}
\newcommand{\gsim}{\mathrel{\mathop{\kern 0pt \rlap
  {\raise.2ex\hbox{$>$}}}
  \lower.9ex\hbox{\kern-.190em $\sim$}}}
\newcommand{\beq}    {\begin{equation}}
\newcommand{\eeq}    {\end{equation}}
\newcommand{\beqarr} {\begin{eqnarray}}
\newcommand{\eeqarr} {\end{eqnarray}}
\newcommand{\barr}   {\begin{array}}
\newcommand{\earr}   {\end{array}}

\begin{document}

\preprint{
\begin{tabular}{r}
DFTT 41/98
\end{tabular}
}

\title{Neutralino properties in the light of a further indication of an 
annual modulation effect in  WIMP direct search}

\author{
\bf A. Bottino$^{\mbox{a}}$
\footnote{E--mail: bottino@to.infn.it,donato@to.infn.it,
fornengo@to.infn.it,scopel@posta.unizar.es},
F. Donato$^{\mbox{a}}$, N. Fornengo$^{\mbox{a}}$, 
S. Scopel$^{\mbox{b}}$\footnote{INFN Post--doctoral Fellow}
\vspace{6mm}
}

\address{
\begin{tabular}{c}
$^{\mbox{a}}$
Dipartimento di Fisica Teorica, Universit\`a di Torino and \\
INFN, Sezione di Torino, Via P. Giuria 1, 10125 Torino, Italy
\\
$^{\mbox{b}}$ Instituto de F\'\i sica Nuclear y Altas Energ\'\i as, \\
Facultad de Ciencias, Universidad de Zaragoza, \\
Plaza de San Francisco s/n, 50009 Zaragoza, Spain
\end{tabular}
}
%\date{August 21, 1998}
\maketitle

\begin{abstract}
We demonstrate that the further indication of a possible annual modulation effect, 
singled out by the DAMA/NaI 
experiment for WIMP direct detection, is widely compatible with an
interpretation in terms of a relic neutralino as the major component of dark
matter in the Universe. We discuss the supersymmetric features of this
neutralino 
in the Minimal Supersymmetric extension of the Standard Model (MSSM) 
and their implications for searches at accelerators. 

\end{abstract}  

\vspace{1cm}


\section{Introduction}

In the seminal papers of Refs.\cite{drukier,freese} it was 
pointed out that the Earth's 
motion around the Sun 
can produce a sizeable 
annual modulation of the signal in experiments of direct search for heavy 
relic particles. 

  Actually, the analysis of a new set of data,  recently collected
by the DAMA/NaI Collaboration (in the period denoted by the Collaboration 
as running period \# 2) \cite{dama2} supports the possible presence of an 
annual 
modulation effect in the counting rate for WIMPs: the hypothesis of
presence of modulation against the hypothesis of absence of  modulation
is statistically favoured at 98.5\% C.L.
Remarkable features of this measurement, obtained with an exposure of 14,962 
kg $\times$ day,  are:

i) An analysis of the experimental data, based on a maximum likelihood method, 
pins down, at a 2--$\sigma$ C.L., a well
delimited region in the plane 
$\xi \sigma^{(\rm nucleon)}_{\rm scalar}$ -- $m_\chi$, 
where $m_\chi$ is the WIMP mass, 
$\sigma^{(\rm nucleon)}_{\rm scalar}$ is the WIMP--nucleon  scalar elastic
cross section  and $\xi = \rho_\chi / \rho_l$ 
is the fractional amount of local 
WIMP density $\rho_\chi$ with respect to the total local 
dark matter density $\rho_l$. This 
$\xi \sigma^{(\rm nucleon)}_{\rm scalar}$ -- $m_\chi$ modulation 
region is shown in Fig. 1, 
which is reproduced here from Fig. 6 of Ref.\cite{dama2} (the values of 
$\xi \sigma^{(\rm nucleon)}_{\rm scalar}$ plotted in Fig. 1 are normalized 
to the value $\rho_l=0.3$ GeV cm$^{-3}$).
The ensuing 1--$\sigma$
ranges for the two quantities are: $m_{\chi} = 59_{- 14}^{+ 22}$ GeV and 
$\xi \sigma^{(\rm nucleon)}_{\rm scalar} = 7.0_{-1.7}^{+0.4} \times 10^{-9}$ nb 
\cite{dama2}. 

ii) The new data confirm a previous indication of an annual modulation (at the
90\% C.L.) found by the same Collaboration, by using a smaller sample of data,
collected in the running period \# 1, with an exposure of 4,549 kg $\times$ day
\cite{dama1}. Most remarkably 
the 2--$\sigma$ C.L. region from data of 
Ref. \cite{dama2} is
entirely contained inside the 90\% C.L. region derived from data of Ref.
\cite{dama1}, also shown in Fig. 1 
(the open solid curve
denotes the 90\% C.L. upper bound derived in Ref.\cite{damapsa}, by 
using pulse shape analysis). 

iii) Because of the property at point ii), 
when the data of the two running periods (with a total exposure of 
19,511 kg $\times$ day) are combined together, one obtains 
 a more delimited 2--$\sigma$ C.L. region  in the plane 
$\xi \sigma^{(\rm nucleon)}_{\rm scalar}$ -- $m_\chi$, 
which is fully embedded in the previous regions. 
Consequently, the determination of 
$m_\chi$ and  $\xi \sigma^{(\rm nucleon)}_{\rm scalar}$ remains very stable: 
$m_{\chi} = 59_{- 14}^{+ 17}$ GeV and 
$\xi \sigma^{(\rm nucleon)}_{\rm scalar} = 
7.0_{-1.2}^{+0.4} \times 10^{-9}$ nb
(if $\rho_l$ is normalized to the value $\rho_l=0.3$ GeV cm$^{-3}$). 
By combining the two sets of data, the
hypothesis of presence of modulation increases to 99.6\% C.L.

It is noticeable  that the two sets of data have been taken under different
operating conditions, since  the experimental set--up was 
dismounted and reassembled between the two running periods.

In extracting the contour lines of Fig. 1 
from the experimental data, the values of some 
astrophysical parameters 
(the root mean square velocity 
$v_{\rm rms}$ of the WIMP 
Maxwellian velocity distribution in the halo,
the WIMP escape velocity $v_{\rm esc}$ in the halo, 
the velocity $v_\odot$ of the Sun around the galactic centre), 
relevant
for the event rates at the detector, had to be chosen.
The values adopted in Fig. 1 
refer to the median values of these parameters in their
experimentally allowed ranges (reported, for instance, in \cite{limiti}), 
namely: 
$v_{\rm rms}$ = 270  Km s$^{-1}$, $v_{\rm esc} = 650$ Km s$^{-1}$,
$v_\odot = 232$ Km s$^{-1}$. 

In Refs. \cite{our1,our2} we derived the theoretical implications of
the experimental data of \cite{dama1}, assuming that the indication of the
possible annual modulation reported there was due to relic
neutralinos. We selected the relevant supersymmetric
configurations and discussed how these may be investigated by
indirect searches for relic WIMPs and at accelerators. 

In the present paper we apply a
similar analysis to the new, much more significant set of data of Ref.
\cite{dama2} and we show that these data are fully compatible with an
interpretation in terms of a relic neutralino as the major component of dark
matter in the Universe. We pin down the regions of the supersymmetric
parameter space relevant for this neutralino and derive the implications for
search at accelerators. 

A  word of caution is in order here. As also remarked in Ref.
\cite{dama2}, although the new DAMA data appear to bring more evidence for a 
possible annual modulation effect, first singled out in Ref.\cite{dama1}, 
this effect awaits further confirmation by additional experimental
investigation in WIMP direct detection \cite{cm}. 
Actually, the DAMA/NaI Collaboration has already collected new data over the
past year; moreover, the experiment still keeps running under good stability
conditions \cite{dama2}
and is expected to provide increasingly significant statistics in
the future. Furthermore, it is remarkable that, as subsequently discussed in
the present paper, the supersymmetric configurations singled out by the
annual modulation effect are also explorable at accelerators and in terms 
of indirect signals of relic neutralinos (i.e., in terms of antiprotons
in space and of up--going muons at neutrino telescopes).

\section{Supersymmetric model}

In this paper we consider the neutralino as a WIMP candidate, able to induce 
annual modulation effects in direct particle dark matter searches. 
This supersymmetric particle is defined 
as the lowest--mass linear superposition of photino ($\tilde \gamma$),
zino ($\tilde Z$) and the two higgsino states
($\tilde H_1^{\circ}$, $\tilde H_2^{\circ}$) \cite{susy}
\begin{equation}
\chi \equiv a_1 \tilde \gamma + a_2 \tilde Z + a_3 \tilde H_1^{\circ}  
+ a_4 \tilde H_2^{\circ}. 
\label{eq:neu}
\end{equation}

We also define a parameter $P \equiv a_1^2 + a_2^2$ in terms of which we
classify neutralinos as: {\it gaugino--like} when $P > 0.9$,
{\it mixed} when $0.1 \leq P \leq 0.9$ and {\it higgsino--like} when $P < 0.1$. 

As a theoretical framework we use the Minimal Supersymmetric extension of the 
Standard Model (MSSM)\cite{susy}, which conveniently  describes the 
supersymmetric phenomenology at the electroweak scale, without too strong 
theoretical assumptions. This model has been extensively used by a number of
authors for evaluations of the neutralino relic abundance and detection rates
 (a list of references may be found, for instance, in  
\cite{our1}). 

The MSSM is based on the same gauge group as the Standard Model, 
contains the supersymmetric extension of its particle content and 
two Higgs doublets $H_1$ and $H_2$. 
As a consequence, the MSSM contains three neutral Higgs fields: two of them 
($h$, $H$) 
are scalar and one ($A)$ is pseudoscalar. 
At the tree level the Higgs sector is specified by two independent parameters:
the mass of one of the physical Higgs fields, which we choose to
be the mass $m_A$ of the neutral pseudoscalar boson, and the ratio of the 
two vacuum expectation values, defined as $\tan\beta\equiv \langle H_2
\rangle/\langle H_1\rangle$. 
Once radiative corrections are introduced, the Higgs sector depends
also on the squark masses through loop diagrams. The radiative corrections 
to the neutral and charged Higgs bosons, employed in the present paper, are 
taken from Refs. \cite{carena,haber}.

The other parameters of the model are defined in the superpotential, 
which contains all the Yukawa interactions
and the Higgs--mixing term 
$\mu H_1 H_2$, and  in the soft--breaking
Lagrangian, which contains the trilinear and bilinear  breaking 
parameters and the soft gaugino and scalar mass terms. 

The MSSM contains a large number of free parameters. 
To cast it into a form adequate for phenomenology, 
it is necessary to introduce a number of 
restrictive assumptions at the electroweak scale.
The usual conditions, which are also employed here, are the following: 
i) all trilinear parameters are set to zero except those of the third family, 
which are unified to a common value $A$;
ii) all squarks and sleptons soft--mass parameters are taken as 
degenerate: $m_{\tilde l_i} = m_{\tilde q_i} \equiv m_0$, 
iii) the gaugino masses are assumed to unify at $M_{GUT}$, and this implies that
the $U(1)$ and $SU(2)$ gaugino masses are related at the electroweak scale by 
$M_1= (5/3) \tan^2 \theta_W M_2$. 

After these conditions are applied, the supersymmetric parameter space
consists of six independent parameters. We choose them to be: 
$M_2, \mu, \tan\beta, m_A, m_0, A$ and vary these parameters in
the following ranges: $10\;\mbox{GeV} \leq M_2 \leq  500\;\mbox{GeV},\;
10\;\mbox{GeV} \leq |\mu| \leq  500\;\mbox{\rm GeV},\;
75\;\mbox{GeV} \leq m_A \leq  1\;\mbox{TeV},\; 
100\;\mbox{GeV} \leq m_0 \leq  1\;\mbox{TeV},\;
-3 \leq A \leq +3,\;
1 \leq \tan \beta \leq 50$. 
We remark that the values taken here as upper limits of the ranges for 
the dimensional parameters, $M_2, \mu, m_0, m_A$, are inspired by the upper 
bounds which may be
derived for these quantities in SUGRA theories, when one requires that the 
electroweak symmetry breaking, radiatively induced by the soft supersymmetry
breaking, does not occur with excessive fine tuning 
(see Ref. \cite{bbefms1} and references quoted therein). 

Our supersymmetric parameter space is further constrained by
all the experimental limits obtained from accelerators on
supersymmetric and Higgs searches. Thus,  the latest data from 
LEP2 on Higgs, neutralino, chargino and sfermion masses are 
used \cite{lep2,ichep}. 

Moreover, the constraints 
due to the $b \rightarrow s + \gamma$ process 
 (see, for instance, 
Refs.\cite{bertolini,bg,garisto,borzumati,wu,barger})  
have to be taken into
account. In our analysis, the inclusive decay rate 
BR($B \rightarrow X_s \gamma$) is calculated with corrections up to the 
leading order. Next--to--leading order corrections 
\cite{chetyrkin,ciuchini1,cza,ciuchini2} 
are included only when 
they can be applied in a consistent way, i.e. both to standard--model 
and  to susy diagrams. This criterion limits the use of 
next--to--leading order corrections to peculiar regions of the supersymmetric 
parameter space, where the assumptions, under which the next--to--leading order 
susy corrections have been obtained, apply \cite{ciuchini2}. 
We require that our theoretical evaluation for BR($B \rightarrow X_s \gamma$) 
is within the  range:  
1.96 $\times 10^{-4} \leq$ BR($B \rightarrow X_s \gamma$) $\leq$ 4.32
$\times 10^{-4}$. This range is obtained by combining the experimental 
data of Refs. \cite{glenn,barate} at 95\% C.L. and by adding a 
theoretical uncertainty of 25\%, whenever the still incomplete 
next--to--leading order susy corrections cannot be applied.

Since we are exploring here the neutralino as a stable dark matter candidate, 
we have to further constrain the
parameter space by requiring that the neutralino is the Lightest 
Supersymmetric Particle (LSP), i.e. we have to exclude regions where the 
gluino or squarks or 
sleptons are lighter than the neutralino. We also have to disregard 
those regions of the parameter space where
the neutralino relic abundance exceeds the cosmological bound, derivable from
measurements of the age of the Universe \cite{age} and 
of the Hubble constant \cite{hubble}. 
Conservatively, for this cosmological bound we take 
$\Omega_{\chi}h^2 \leq 0.7$ 
($h$ is the usual Hubble parameter, defined in terms of the present--day 
value $H_0$ of the Hubble constant as 
$h \equiv H_0/(100~$ km$~$s$^{-1}~$Mpc$^{-1})$). 
The neutralino relic abundance is calculated here as illustrated in
Ref.\cite{ouromega}. 
Inclusion of 
coannihilation effects \cite{coannih}
in the calculation of $\Omega_{\chi} h^2$ are not necessary here, 
since the instances under which these effects might be sizeable are 
marginal in our supersymmetric parameter space.

\section{Selection of supersymmetric configurations by the annual modulation
effect}

We discuss now which region in the susy parameter space is selected by the 
new DAMA modulation data \cite{dama2}.

Let us start by converting the region  delimited by the 2--$\sigma$ 
C.L. dashed  contour line
in the plane $\xi \sigma^{(\rm nucleon)}_{\rm scalar}$ -- $m_\chi$ of Fig. 1 
into an enlarged one, which accounts for the uncertainty in the value of 
$\rho_l$. If  a possible flattening of the dark matter
halo  \cite{turner1}  and a possibly sizeable baryonic contribution to the
galactic dark matter \cite{turner} are taken into account, the following range 
for $\rho_l$ has conservatively to be taken:
0.1 GeV cm$^{-3} \leq \rho_l \leq $ 0.7 GeV cm$^{-3}$. One then obtains from the
2--$\sigma$ C.L. region of Fig. 1, where the total dark matter density is 
normalized to
the value $\rho_l=0.3$ GeV cm$^{-3}$, the relevant 2--$\sigma$ C.L. 
region of Fig. 2 (hereafter denoted as region $R$).

Now we have to find which supersymmetric configurations, out of those in the 
parameter space outlined in Sect. II, are selected by the requirement that 
($m_{\chi}, \xi \sigma_{\rm scalar}^{(\rm nucleon)}) \in R$. 
To this purpose we evaluate 
$m_{\chi}$,  $\sigma_{\rm scalar}^{(\rm nucleon)}$ and $\xi$ in the MSSM scheme 
previously defined.

The neutralino mass is evaluated  as usual by taking the lowest mass 
eigenstate of the neutralino mass matrix \cite{susy}.

The neutralino--nucleon scalar cross--section is calculated with the formula

\beq
\sigma_{\rm scalar}^{(\rm nucleon)} = \frac {8 G_F^2} {\pi} M_Z^2 m_{\rm red}^2 
\left[\frac{F_h I_h}{m_h^2}+\frac{F_H I_H}{m_H^2}+
\frac{M_Z}{2} \sum_q <N|\bar{q}q|N>
\sum_i P_{\bar q_i} ( A_{\tilde{q}_i}^2- B_{\tilde{q}_i}^2) \right]^2, 
\label{eq:sigma}
\eeq

\noindent
where the two first terms inside the brackets refer to the diagrams with 
$h$-- and $H$--exchanges in the t--channel 
(the $A$--exchange diagram is strongly kinematically suppressed and then omitted
here) \cite{barbieri} and the third term refers 
to the graphs with squark--exchanges in the s-- and u--channels \cite{griest}.
The mass $m_{\rm red}$ is the neutralino--nucleon reduced mass and 

\beqarr
F_h &=& (-a_1 \sin \theta_W+a_2 \cos \theta_W) 
          (a_3 \sin \alpha + a_4 \cos \alpha) 
           \nonumber \\
           F_H &=& (-a_1 \sin \theta_W+a_2 \cos \theta_W) 
           (a_3 \cos \alpha - a_4 \sin \alpha) \nonumber \\
           I_{h,H}&=&\sum_q k_q^{h,H} m_q \langle N|\bar{q} q |N \rangle.
\label{effe}
\eeqarr

The angle $\alpha$ rotates $H_1^{(0)}$ and $H_2^{(0)}$ into $h$ and $H$, and 
the coefficients $k_q^{h,H}$ are given by 
$k_{u{\rm -type}}^h = $cos$\alpha / $sin$\beta$ and  
$k_{u{\rm -type}}^H = - $sin$\alpha /$ sin$\beta$ for the up--type quarks, 
and by $k_{d{\rm -type}}^h = - $sin$\alpha / $cos$\beta$ and 
$k_{d{\rm -type}}^H = -$ cos$\alpha / $cos$\beta$ for the down--type quarks. 
The matrix elements $<N|\bar q q|N>$ are meant over the nucleonic state. 
By using the heavy quark expansion \cite{svz}, one may rewrite the quantity
$I_{h,H}$ as follows 

\beq 
I_{h,H} = k_{u{\rm -type}}^{h,H} g_u + k_{d{\rm -type}}^{h,H} g_d,
\label{eq:i}
\eeq

\noindent
where

\beq
g_u = \frac {4} {27} (m_N + \frac {19}{8} \sigma_{\pi N} 
- a \sigma_{\pi N}),~~~~~g_d = \frac{2}{27} (m_N + \frac{23}{4} \sigma_{\pi N} 
+ \frac{25}{2} a \sigma_{\pi N}).
\eeq
\noindent
Here $\sigma_{\pi N}$ is the so--called pion--nucleon sigma term, 
$\sigma_{\pi N} = \frac{1}{2} (m_u + m_d) <N|\bar uu + \bar dd|N>$, and the
parameter $a$ is related to the strange--quark content of the nucleon $y$ by 

\beq
a = y \frac{m_s}{m_u + m_d},~~~~y=2 \frac{<N|\bar ss|N>}
{<N|\bar uu+ \bar dd|N>}. 
\eeq
\noindent
For these parameters we use the following values: 
$\sigma_{\pi N} = 45$ GeV \cite{gasser}, $y = 0.33 \pm 0.09$ \cite{liu} and 
2$m_s/(m_u + m_d) = 29$ \cite{bj}; thus, using the central value of $y$, 
we obtain $g_u = 123$ GeV and $g_d = 288$ GeV. 

In the squark--exchange terms of Eq.(\ref{eq:sigma})  $\sum_i$ denotes a sum 
over the mass eigenstates, $P_{\tilde q_i}$ stands for the squark propagators 

\beq
P_{\tilde{q}_i}=\frac{1}{2}\left(
\frac{1}{ m_{\tilde{q}_i}^2-(m_{\chi}-m_q)^2}+
\frac{1}{m_{\tilde{q}_i}^2-(m_{\chi}+m_q)^2}
\right),
\label{eq:prop}
\eeq
\noindent 
\noindent
and the $A_{\tilde q_i}$ and $B_{\tilde q_i}$ coefficients are given by 

\beqarr
A_{\tilde{q_1}}&=&\cos \theta_q (X_q+Z_q)+\sin \theta_q (Y_q+Z_q) \nonumber \\
B_{\tilde{q_1}}&=&\cos \theta_q (X_q-Z_q)+\sin \theta_q (Z_q-Y_q) \nonumber \\
X_q&=&-\left( \cos \theta_W T_{3q}a_2+\sin \theta_W \frac{Y_{qL}}{2}a_1 \right) 
\;\;\;\; ; \;\;\;\; Y_q\;=\;\sin \theta_W \frac{Y_{qR}}{2}a_1 \nonumber \\
%Y_q&=&\sin \theta_W \frac{Y_{qR}}{2}a_1 \nonumber \\
Z_{u{\rm -type}}&=&-\frac{m_{u{\rm -type}} a_4}{2 \sin\beta M_Z} 
\;\;\;\; ; \;\;\;\; 
Z_{d{\rm -type}}=-\frac{m_{d{\rm -type}} a_3}{2 \cos\beta M_Z},
\eeqarr
where $T_{3q}$, $Y_{qL}$, $Y_{qR}$ refer to the isospin and to the 
hypercharge quantum 
numbers of $\tilde{q}_{L,R}$, 
respectively.
The couplings $A_{\tilde{q_2}}$ and $B_{\tilde{q_2}}$ may be obtained with 
the substitution
$\sin \theta_q$ $\rightarrow$ $\cos \theta_q$ and
$\cos \theta_q$ $\rightarrow$ $-\sin \theta_q$.

In our numerical applications the squark propagators in Eq.(\ref{eq:prop}) 
have been regularized by inserting appropriate widths in the denominators. 
In general, it turns out that the Higgs--exchange 
amplitudes are largely dominant over the squark--exchange ones, the
latter  competing with the former ones almost exclusively when an 
enhancement
in their size is originated by a mass fine--tuning in the squark--propagator 
denominators.

As for the values to be assigned to the quantity $\xi = \rho_{\chi}/ \rho_l$ 
we adopt the standard rescaling recipe \cite{gaisser}. 
For each point of the parameter
space, we take into account the relevant value of the cosmological neutralino
relic density. When $\Omega_\chi h^2$ is larger than a minimal value
$(\Omega h^2)_{\rm min}$, compatible with observational data and with 
large--scale 
structure calculations, we simply put $\xi=1$.
When $\Omega_\chi h^2$ turns out  to be less than $(\Omega h^2)_{\rm min}$, 
and then the neutralino may only provide a fractional contribution
${\Omega_\chi h^2 / (\Omega h^2)_{\rm min}}$
to $\Omega h^2$, we take $\xi = {\Omega_\chi h^2 / (\Omega h^2)_{\rm min}}$.
The value to be assigned to $(\Omega h^2)_{\rm min}$ is
somewhat arbitrary, in the range 
$0.01 \lsim (\Omega h^2)_{\rm min} \lsim 0.3$. We use here the value 
$(\Omega h^2)_{\rm min} = 0.01$, which is conservatively derived from the
estimate $\Omega_{\rm galactic} \sim 0.03$.

Using the previous formulae we find that a large portion of the 
modulation region $R$ is indeed covered by supersymmetric configurations, 
compatible with all present physical constraints. This set of susy states, 
which will hereafter be denoted as set $S$, is displayed in Fig. 2 with 
different symbols, depending on the neutralino composition. 
In Fig. 2(a) we notice that a quite sizeable portion of region $R$ is 
populated by supersymmetric configurations
with neutralino relic abundance inside the cosmologically interesting range 
$0.01 \lsim \Omega_{\chi} h^2 \lsim 0.7$. 
Thus we obtain  the first main result of our analysis, i.e. 
{\it the annual 
modulation region, singled out by the DAMA/NaI experiment, is largely 
compatible with a relic neutralino as the major component of dark matter}. 
This is certainly the most remarkable possibility. However, we also keep under
consideration neutralino configurations with a small contribution to 
$\Omega_{\chi} h^2$ (see Fig. 2(b)), since also the detection of relic 
particles with these features would provide in itself a very noticeable 
information.

The neutralino relic abundance $\Omega_{\chi} h^2$ is plotted versus the 
quantity 
$\xi \sigma_{\rm scalar}^{(\rm nucleon)}$ in terms of the neutralino 
composition in 
Fig. 3. Here we remark some anticorrelation between the two plotted quantities. 
This feature is expected on general grounds, as discussed for instance in 
Ref. \cite{bbefms}. In fact, it is due to the combination of two properties: 
(i) the direct detection rate is proportional to 
$\sigma_{\rm scalar}^{(\rm nucleon)}$, 
and $\Omega_{\chi} h^2 \propto \sigma_{\rm ann}^{-1}$, 
where $\sigma_{\rm ann}$ is the neutralino--neutralino 
annihilation cross--section, (ii) usually $\sigma_{\rm ann}$ 
and $\sigma_{\rm scalar}^{(\rm nucleon)}$, as functions of the supersymmetric 
model parameters, are
either both increasing or both decreasing. Therefore, neutralinos with lower
values for the relic abundance have higher couplings with matter (this feature
is attenuated, when rescaling in $\rho_{\chi}$ is operative; this 
occurs  here
for $\Omega_{\chi} h^2 < 0.01$). 

In view of the discussed anticorrelation between 
$\sigma_{\rm scalar}^{(\rm nucleon)}$ and $\Omega_{\chi} h^2$, 
it is remarkable that the relatively large neutralino--matter cross--sections,
implied by the DAMA modulation effect, agree with a relic
neutralino making up a major  contribution to dark matter, i.e. 
with a neutralino whose relic abundance falls into the cosmologically 
interesting range $0.01 \lsim \Omega_{\chi} h^2 \lsim 0.7$. 
Most of the neutralino configurations falling in this range of $\Omega_{\chi}
h^2$ turn out to be gaugino--like.

We further notice that recent observations and
analyses \cite{omegamatter} point to values of 
$\Omega_{\rm matter}$ somewhat smaller than those considered in the past: 
$0.1 \lsim \Omega_{\rm matter} \lsim 0.4$. 
If we combine this range with the one for $h$: 
$0.55 \lsim h \lsim 0.80$ \cite{hubble} and require that a 
cold dark matter candidate (such as the neutralino) supplies
$\sim (80$--$90)\%$ of $\Omega_{\rm matter}$, we obtain: 
$0.02 \lsim \Omega_{\rm CDM} h^2 \lsim 0.2$. This turns out to be the most
appealing interval for relic neutralinos. It is remarkable that this range for 
$\Omega_{\chi} h^2$ is densely populated by configurations of set $S$ (see 
Fig. 3).

\section{Further properties of the configurations singled out by the 
annual modulation effect} 

Let us proceed  now to an analysis of other 
 main properties of the configurations of set $S$, related to a possible
 investigation of these supersymmetric states  at accelerators.

As is already clear from Fig. 2, the set $S$ contains neutralino 
 compositions of various nature, from higgsino--like  to gaugino--like ones. 
This property is further displayed in 
Fig. 4, where we show the location of the configurations of set $S$ in the plane
$\mu$--$M_2$, for two representative values of $\tan \beta$.

The properties of our set $S$ relevant to searches of neutral Higgses at
accelerators are displayed in Fig. 5. Section (a) of this figure 
shows a scatter plot of set $S$ in term  of $m_h$ and  $\tan \beta$, 
section (b) provides essentially the same information, but in terms of 
$m_h$ and the quantity $\sin^2 (\alpha - \beta)$, which is the relevant 
coupling for the 
channels of possible neutral Higgs production at LEP. In the plot of section
(a) it is apparent  a correlation between $\tan \beta$ 
and $m_h$. This is due to the fact that the rather 
large  values of the neutralino--nucleon scalar cross--section, 
$\sigma_{\rm scalar}^{(\rm nucleon)} \sim (10^{-9} - 10^{-8}$) nb, as required
 by the annual modulation data, impose that either the couplings are large 
(then large $\tan \beta$) and/or the
process goes through the exchange of  a light particle. 
Thus, Higgs--exchange dominance and 
$\sigma_{\rm scalar}^{(\rm nucleon)} \sim (10^{-9} - 10^{-8})$ nb require a 
very light $h$ at small $\tan \beta$, 
and even  put {\it a lower bound on tan $\beta$: $\tan \beta \gsim 2.5$}. 
At larger values of $\tan \beta$, the mass $m_h$ is less constrained, also 
because, at large $\tan \beta$,  the squark--exchange diagrams 
may occasionally compete with the Higgs--exchange ones in keeping 
$\xi \sigma_{\rm scalar}^{(\rm nucleon)}$ at a sizeable value. 
{From} Fig. 5(a) we notice that a good deal of susy
configurations are explorable at LEP2, while others will require experimental
investigation at a high luminosity Fermilab Tevatron, which 
should be capable to explore Higgs masses up to $m_h \sim$ 130 GeV 
\cite{tev,baer}.

In Fig. 6 the  configurations of set $S$ are shown in the plane
$m_{\tilde t_1} - \tan \beta$ ($t_{\tilde t_1}$ denotes the lightest 
top--squark).
 This scatter plot reveals  an interesting correlation:
 at small $\tan \beta$ only light $\tilde t_1$'s are allowed. 
In the Appendix it is shown that this feature occurs as a joint effect due to 
the $b \rightarrow s + \gamma$ constraint and to the annual modulation data 
\cite{bsg}.

{From} the previous results, it then turns out that annual modulation data and 
$b \rightarrow s + \gamma$ constraint complement each other in providing 
stringent bounds on both $m_h$ and $m_{\tilde t_1}$, at small $\tan \beta$. For
instance, for $\tan \beta \lsim 5$ one has $m_h \lsim $ 105 GeV and  
$m_{\tilde t_1} \lsim$ 350 GeV.  

Finally, in Fig. 7 we display the scatter plot of set $S$ in the plane 
$m_{\chi}$ -- $\tan \beta$. Since the reach of LEP2 extends only up to
the dashed vertical line, at $m_\chi \simeq 50$ GeV, the exploration of the 
whole
interesting region will require Tevatron upgrades or LHC. Under favorable
hypothesis, TeV33 could provide exploration up to the vertical solid line.

Apart from exploration at accelerators, configurations of set $S$ may be
investigated by means of indirect measurements of relic neutralinos, 
such as cosmic--ray antiprotons \cite{pierre} and
 neutrino fluxes from Earth and Sun (\cite{bbefms} and references quoted
therein). 
On the basis of a preliminary analysis, we found that configurations of set 
$S$ provide quite significant signals in both instances. In the case of 
antiprotons, a large
fraction of configurations of set $S$ provide $\bar p$ fluxes at the
level of the measurements by the balloon--borne BESS experiment 
 \cite{bess}.
These configurations will be further investigated with the data 
collected during the Shuttle flight by the AMS experiment \cite{ams}.
A similar situation occurs for
the neutrino fluxes induced by configurations of set $S$, which turn out
to be within the reach of MACRO \cite{macro} and Baksan \cite {baksan} 
neutrino telescopes.
Details of our analysis on the indirect detection searches are 
presented in Ref.\cite{indirectnew}.

We end this section by some more general theoretical considerations. 
We have discussed here  the physical 
implications of the annual modulation data in the framework of a 
 MSSM  at the electroweak scale, since this scheme provides the 
simplest and least--constrained model for discussing susy phenomenology.  
 However, we have also performed an 
analysis of the modulation data in the framework of Supergravity 
(SUGRA) theories. The results of this study are presented in 
Ref. \cite{companion}. We simply report here  that we have 
ascertained   that a fraction of configurations of set $S$ are indeed 
compatible with SUGRA schemes, even more so when the unification 
conditions, which are usually imposed at GUT scale, are 
somewhat relaxed, 
 for instance by allowing deviations from a strict 
unification assumption in the Higgs masses at the GUT scale 
\cite{bbefms1}.  It is remarkable that these configurations fall into the 
region of susy parameter space where electroweak symmetry breaking 
occurs without excessive fine tuning between competing terms. A simple 
case of this feature occurs for the neutralino mass, whose range for the 
annual modulation configurations is well within the 
no--fine--tuning upper bound $m_{\chi} \lsim $ O(100 GeV) \cite{bbefms1}.

\section{Conclusions}

The new data of the 
DAMA/NaI experiment \cite{dama2}, which support a possible annual 
modulation effect in 
the counting rates for relic WIMPs, previously reported by the same
Collaboration \cite{dama1}, have been analysed here in terms of relic 
neutralinos. {\it We have proved that the annual modulation data are largely 
compatible with a relic 
neutralino making up the major part of dark matter in the Universe}.

We have also investigated the possibility of exploring the supersymmetric
states, selected by the annual modulation data, at accelerators.
We have demonstrated that an analysis of the main features of these 
susy configurations 
is within the reach of present or planned experimental set--ups. 
In particular, we have found the following results:

a) The sizeable neutralino--nucleon elastic cross--sections, implied by the 
annual modulation data, entail  a rather stringent upper bound for 
$m_h$ in terms of $\tan \beta$. In particular, this property  implies that no 
susy configuration would be allowed for $\tan \beta \lsim 2.5$. 
A large portion of the region covered by the scatter plot in the plane 
$m_h$ -- $\tan \beta$ is explorable at LEP2, the remaining one will be at 
TeV33.

b) The annual modulation data and the $b \rightarrow s + \gamma$ 
constraint complement each other in providing a correlation 
between $\tan \beta$ and the mass of the lightest top--squark. 

As remarked in the introduction, a solid confirmation of the
annual modulation effect as singled out by the DAMA/NaI Collaboration 
will require further accumulation of an increasingly significant 
statistics with  very stable set--ups over a few years. However, 
it is worth noticing that the detection of this effect, 
if confirmed by further experimental evidence, would 
turn out to be a major breakthrough in establishing the existence of 
particle dark matter in the Universe. It is very rewarding that the 
features of this dark matter particle are widely compatible with those 
expected for the neutralino, both in MSSM and 
in SUGRA schemes, and that several of its 
properties can be explored 
in the near future at accelerators and by indirect searches for
relic neutralinos.

\vspace{1cm}
\section { Acknowledgements}

We wish to thank Prof. R. Bernabei and Dr. P. Belli for very
interesting discussions about experimental aspects of the DAMA/NaI 
experiment and about the analysis of their data. We also thank Dr. P. Gambino 
for informative discussions on the next--to--leading order corrections to 
$b \rightarrow  s + \gamma$.

\vspace{1cm}
\section {Appendix}

Here we discuss the origin of the correlation between $m_{\tilde t_1}$ and 
$\tan \beta$ which is apparent in the plot of Fig. 6  at small $\tan \beta$. 

 Let us start by considering how the $b \rightarrow s + \gamma$ 
constraint 
\cite{bertolini,bg,garisto,borzumati,wu,barger} 
correlates the three parameters $\tan \beta, m_h$ and 
$m_{\tilde t_1}$. Thus, leaving momentarily aside the annual modulation data, 
let us consider in the plane 
$m_{\tilde t_1}$--$\tan \beta$ the regions of our parameter space 
which satisfy all accelerator constraints (including $b \rightarrow s +
\gamma$) and the further requirement that 
$m_h$ is below some arbitrarily fixed value $m_h^*$. 
In Fig. A.1 these regions are represented by the domains on the left
of the various lines, which are denoted by the following values of 
$m_h^*$: $m_h^* = 80, 90, 100, 110$ GeV.
It is possible to show that the $b \rightarrow s + \gamma$ constraint is
instrumental in establishing the peculiar shape of the various contour lines 
at fixed $m_h$. 

If we now combine the plot of Fig. A.1 with the one of Fig. 5(a) we obtain
the situation displayed in Fig. A.2, the allowed region being the 
one on the left of the various curves, depending on the values of 
$m_h^*$.  {From} this figure we see how the 
$\tan \beta$--$m_{\tilde t_1}$ correlation, occurring in Fig. 6, 
 is due to the joint effect of 
$b \rightarrow s + \gamma$  and annual modulation data.

\vfill
\eject

\begin{center}
{\large FIGURE CAPTIONS}
\end{center}
\vspace{1cm}

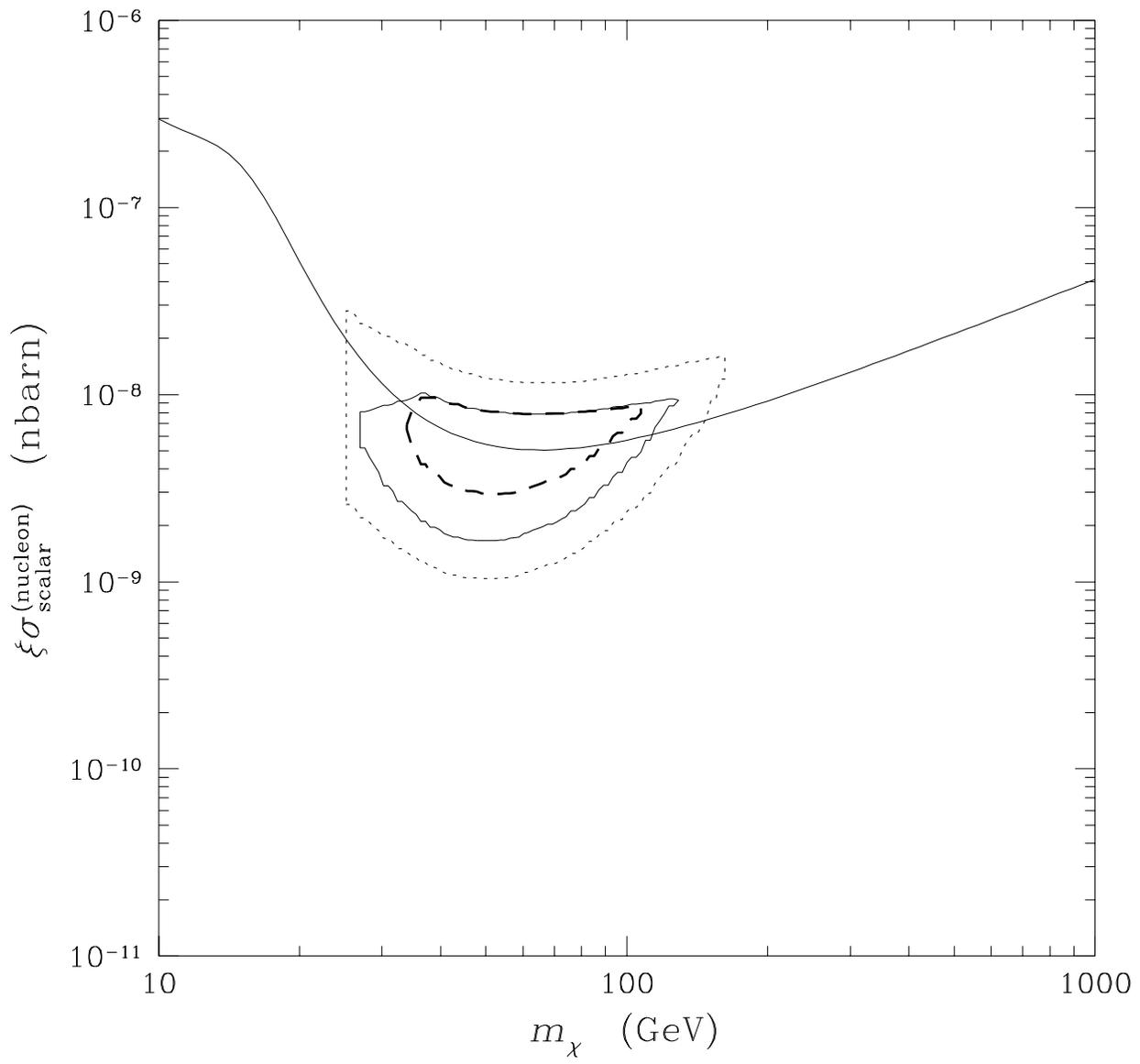
{\bf Figure 1} -- 
Annual modulation regions singled out by the DAMA/NaI experiments in the plane 
$m_{\chi}$--$\xi \sigma_{\rm scalar}^{(\rm nucleon)}$. The dotted contour line 
denotes the 90\% C.L. region deduced from the data of the running period 
\# 1\cite{dama1},
the solid contour line delimits the 2--$\sigma$ C.L. region deduced from 
the data of the running period \# 2 \cite{dama2}, and the dashed contour line
delimits the 2--$\sigma$ C.L. region, obtained by combining together the data of
the two running periods. The solid open curve denotes the 90\% C.L. upper
bound, obtained in Ref.\cite{damapsa}, 
where a pulse shape analysis of the events was used. 
This figure is reproduced from Fig. 6 of Ref.\cite{dama2}.

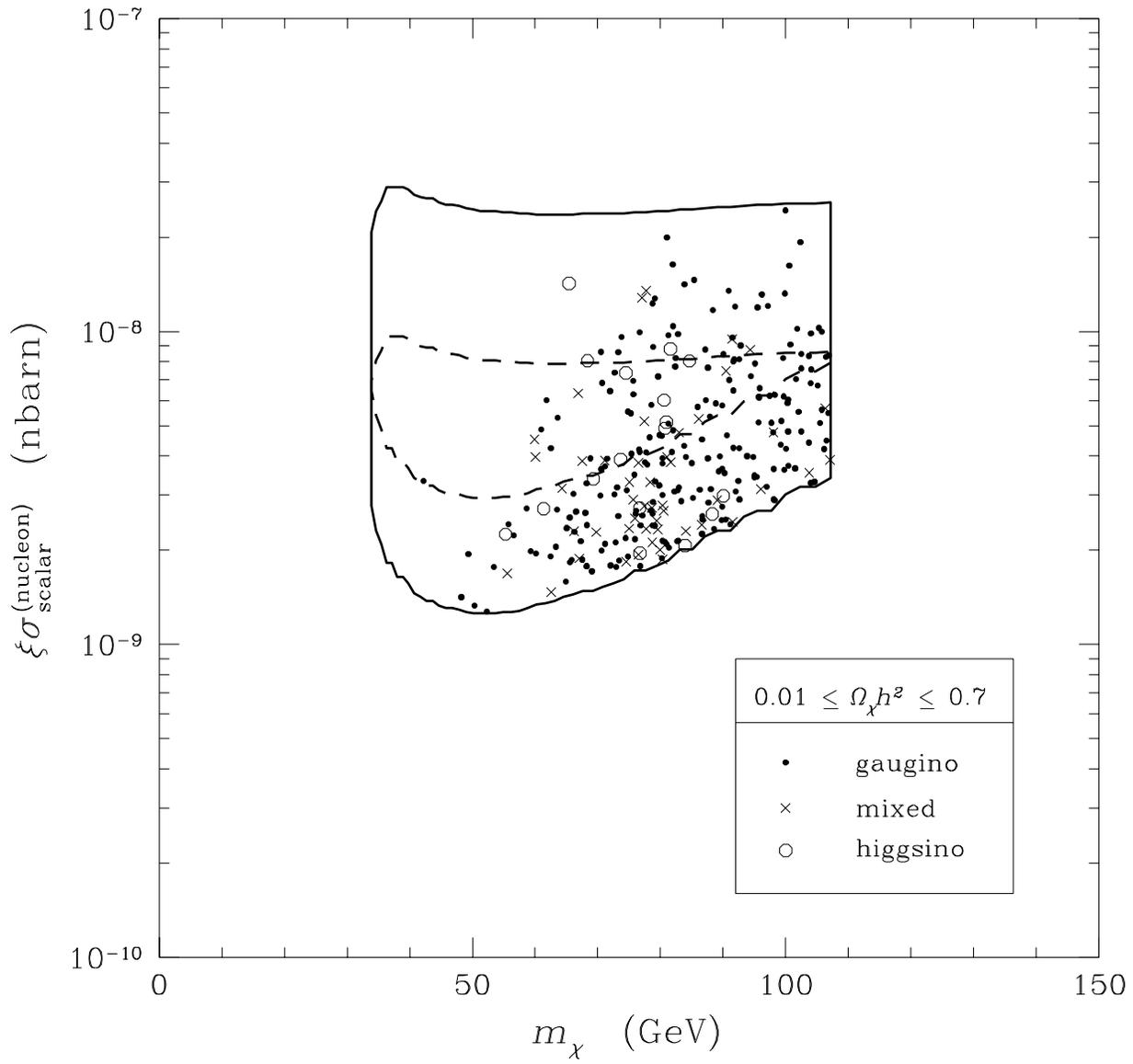
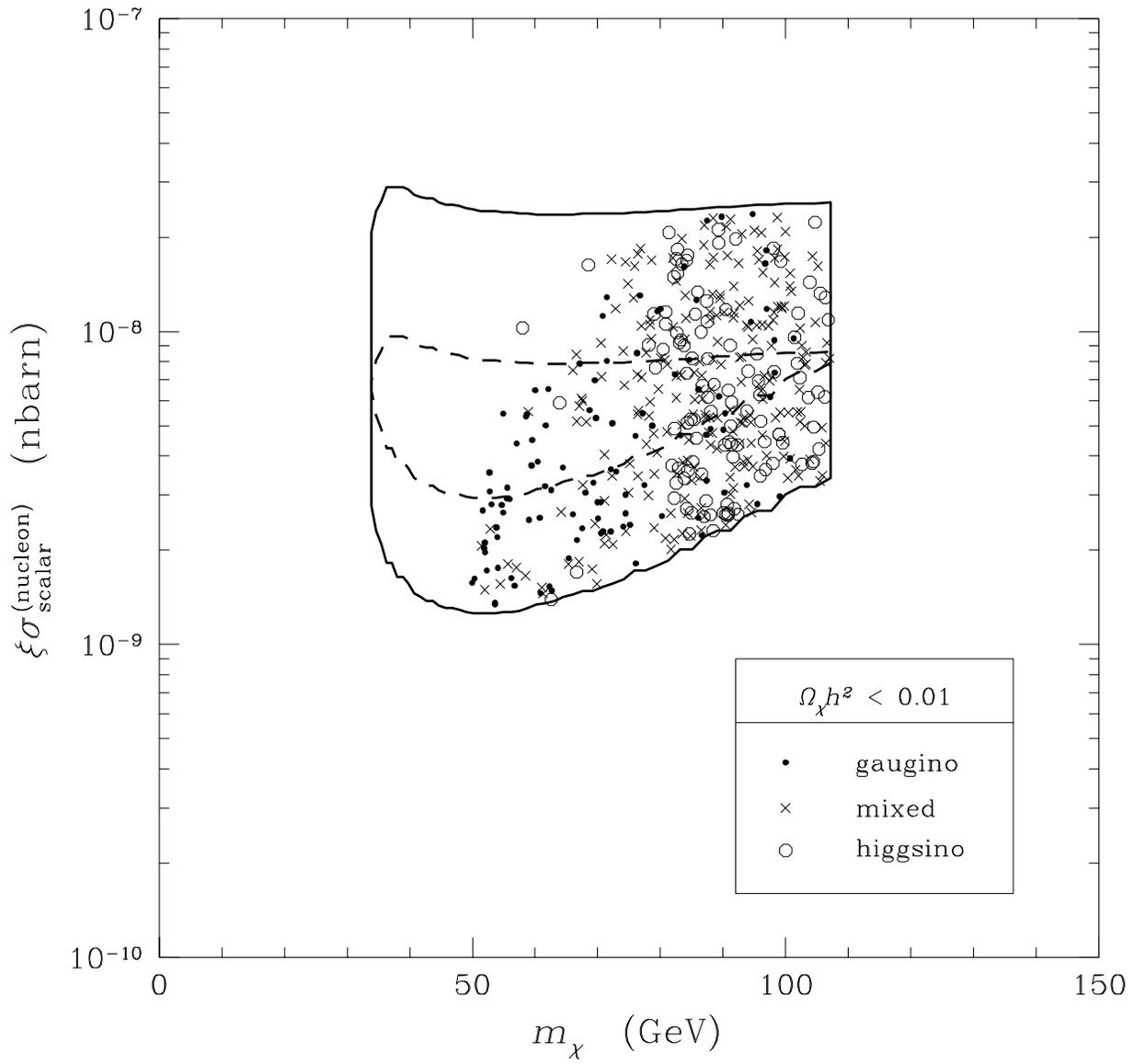
{\bf Figure 2} -- Scatter plot of set $S$ in the plane 
$m_{\chi}$--$\xi \sigma_{\rm scalar}^{(\rm nucleon)}$. 
The dashed contour line
delimits the 2--$\sigma$ C.L. region, obtained by the DAMA/NaI Collaboration, 
 by combining together the data of
the two running periods of the annual modulation experiment \cite{dama2}.  
The solid contour  line is obtained from the dashed line, which refers to the
value $\rho_l = 0.3 ~$ GeV cm$^{-3}$, by accounting for 
the uncertainty range of $\rho_l$, as explained in Sect. III (the region 
delimited by the solid line is denoted as region $R$ in the text). 
Displayed in this figure are only the representative points of the susy 
parameter space, defined in Sect. II, which fall inside the region $R$. 
 Dots, crosses and circles denote neutralino compositions
according to the classification given in Sect. II.
Sections (a) and (b) refer to configurations with 
$0.01 \leq \Omega_{\chi} h^2 \leq 0.7$ and with $\Omega_{\chi} h^2 < 0.01$, 
respectively. 

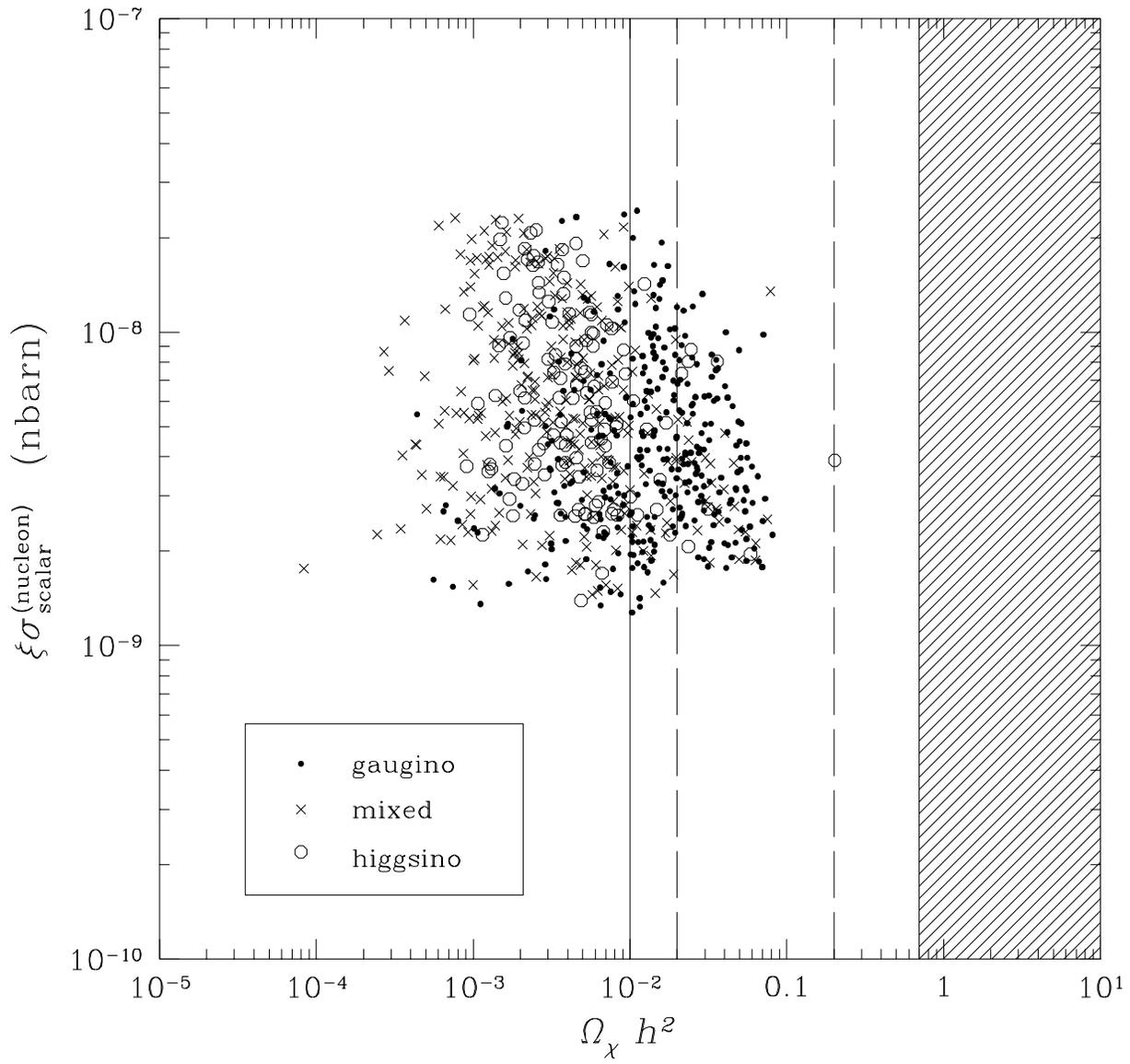
{\bf Figure 3} -- Scatter plot of set $S$ in the plane 
 $\Omega_{\chi} h^2$ -- $\xi \sigma_{\rm scalar}^{(\rm nucleon)}$.
Dots, crosses and circles denote neutralino compositions
according to the classification given in Sect. II. The two vertical solid lines 
delimit the $\Omega_{\chi} h^2$--range of cosmological interest. The two
dashed lines delimit the most appealing interval for 
$\Omega_{\chi} h^2$, as suggested by the most recent observational data.

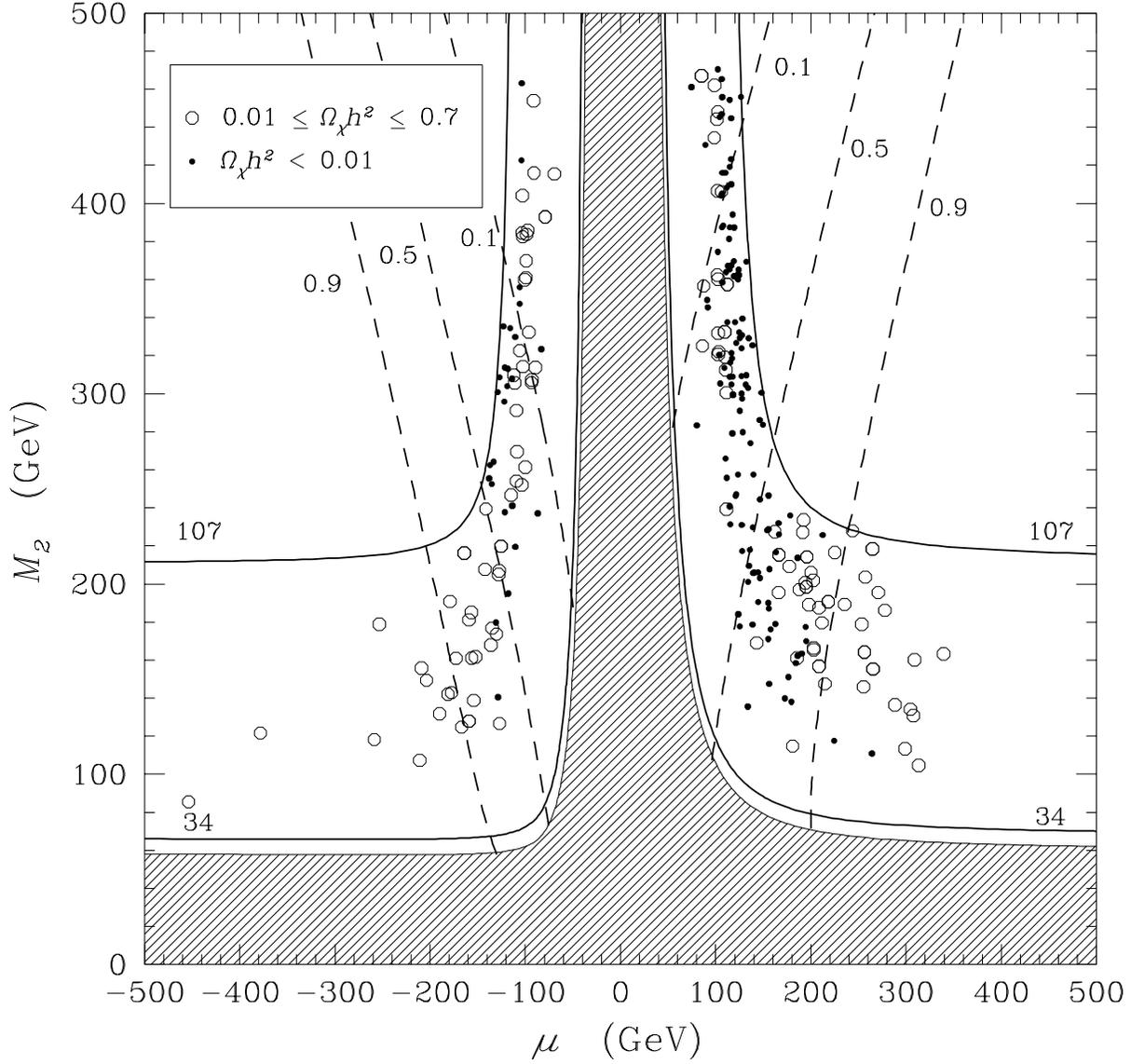
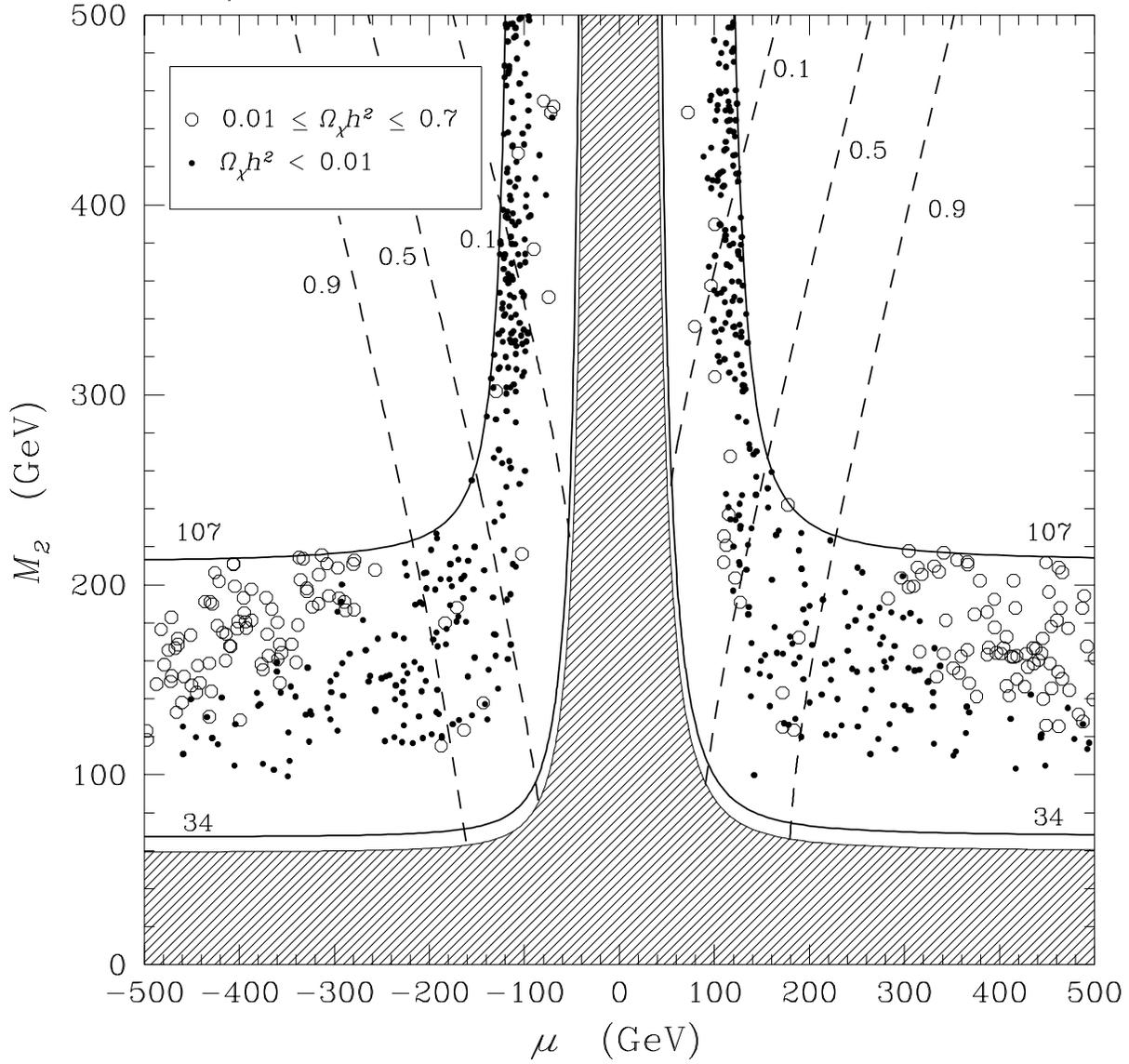
{\bf Figure 4} -- Scatter plot of set $S$ in the plane 
$\mu$--$M_2$. Sections (a) and (b) 
refer to two representative values of  $\tan \beta$: 
$\tan \beta = 8$ and $\tan \beta = 30$, respectively. 
The solid curves denote the iso--mass curves which delimit the annual
modulation region $R$, i.e. the iso--mass curves for $m_{\chi}$ = 34 GeV and 
$m_{\chi}$ = 107 GeV. The dashed curves denote the neutralino composition, 
and correspond to $P$ = 0.1, 0.5, 0.9. The hatched region is excluded by LEP at
$\sqrt s$ = 183 GeV.  

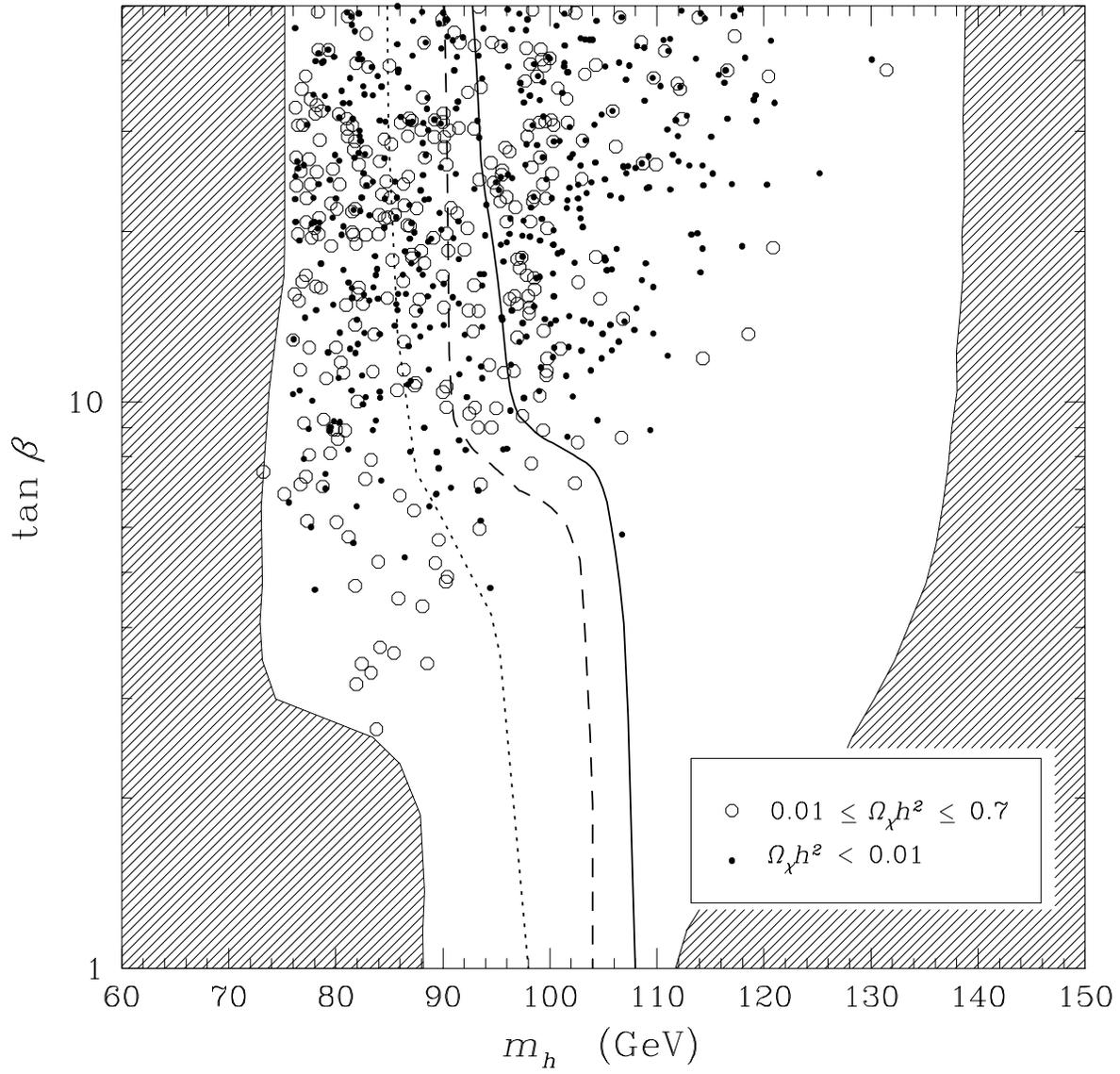
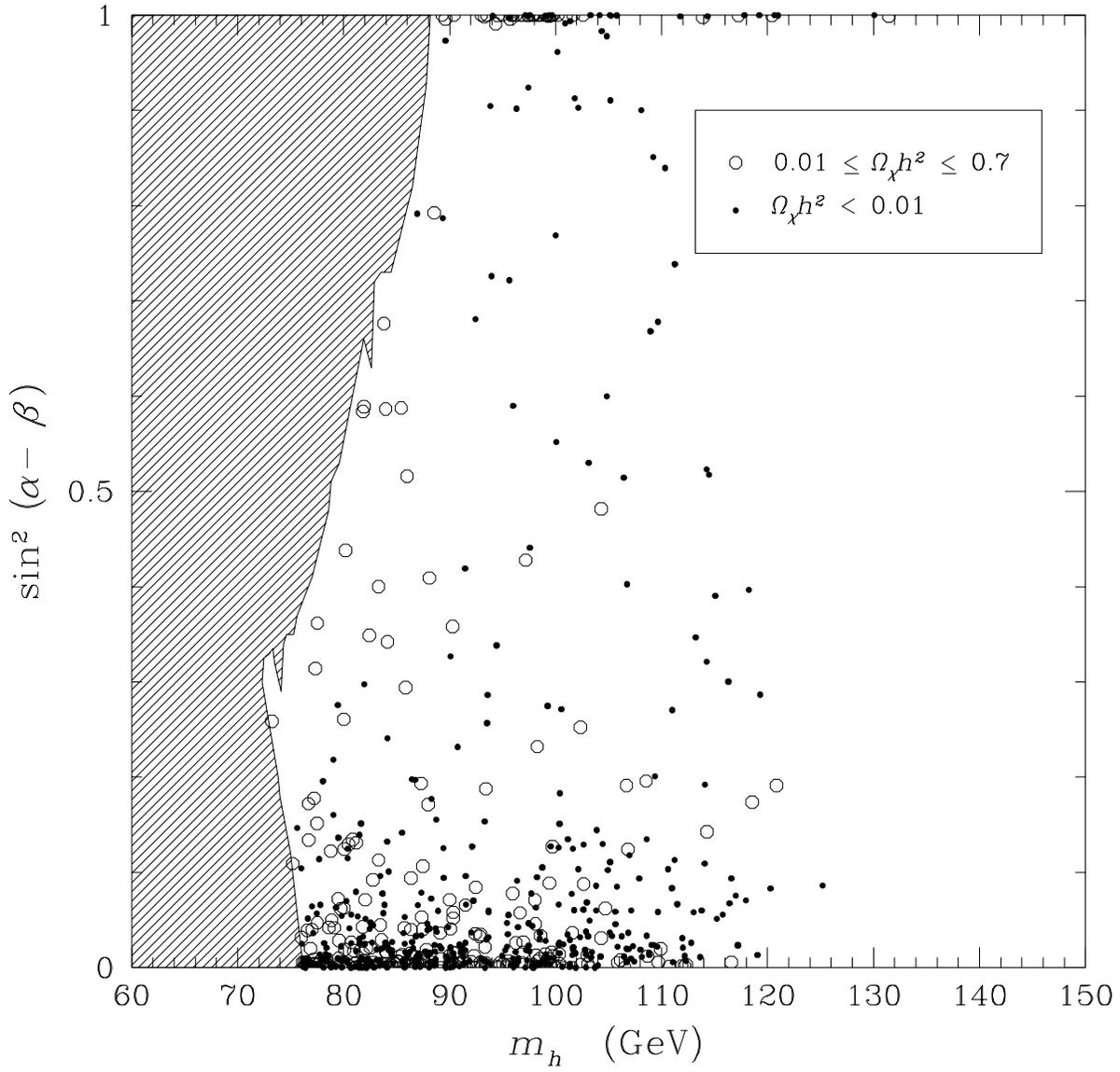
{\bf Figure 5} -- Section (a) --
Scatter plot for set $S$ in the plane $m_h$ -- $\tan \beta$. 
The hatched region on the right is excluded by theory. 
The hatched region on the left is 
excluded by present LEP data at $\sqrt s$ = 183 GeV. The dotted and the dashed 
curves denote the reach of LEP2 at energies $\sqrt s$ = 192 GeV and 
$\sqrt s$ = 200 GeV, respectively. The solid line represents the 
95\% C.L. bound reachable at LEP2, in case of non discovery of a neutral 
Higgs boson. \\
Section (b) --
Scatter plot for set $S$ in the plane $m_h$ -- $\sin^2 (\alpha - \beta)$. 
The hatched region on the left is 
excluded by LEP data at $\sqrt s$ = 183 GeV.

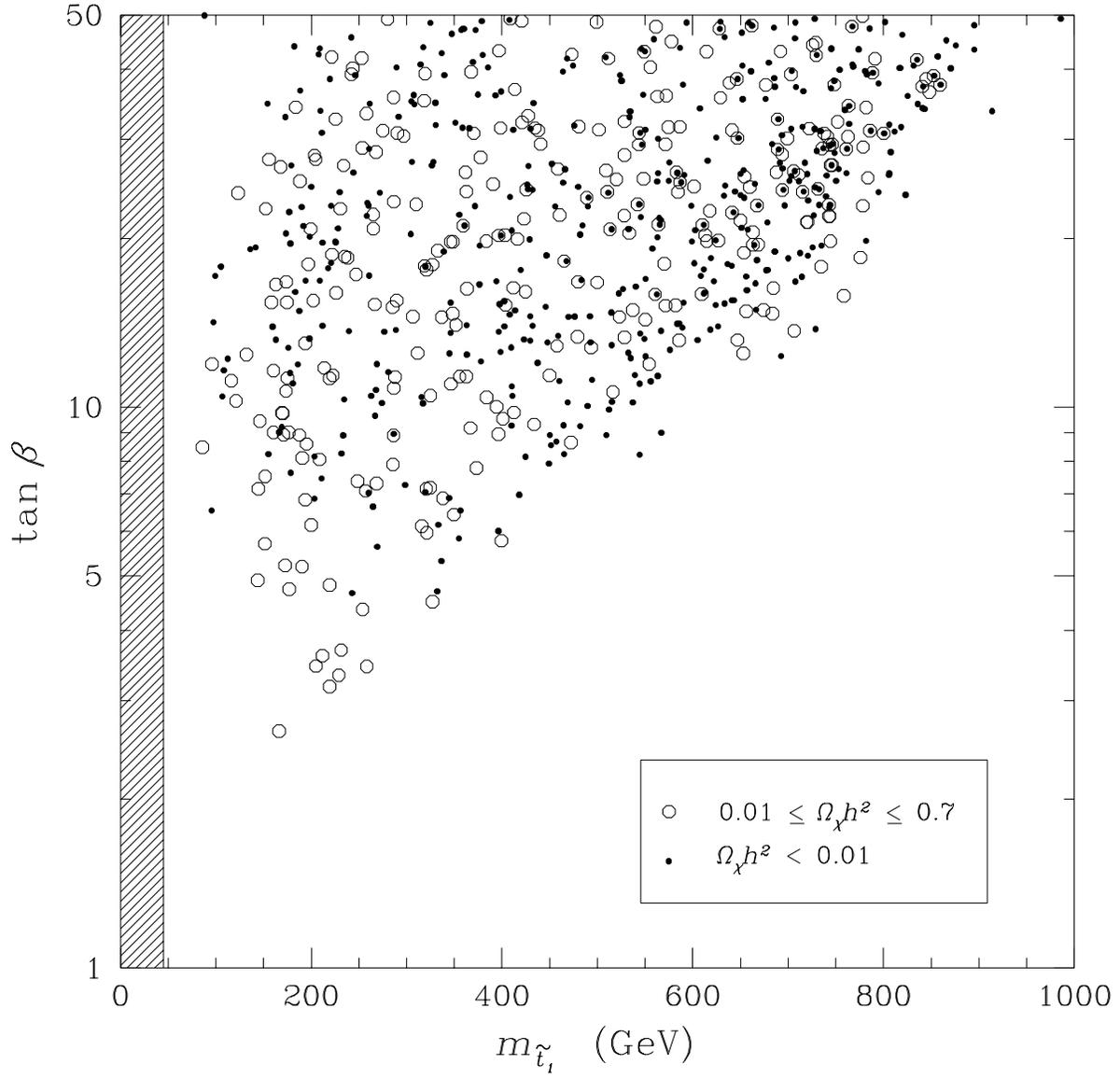
{\bf Figure 6} -- Scatter plot for set $S$ in the plane 
$m_{\tilde t_1}$ -- $\tan \beta$.  
The hatched region is excluded by LEP data (without any restriction on other 
masses). 

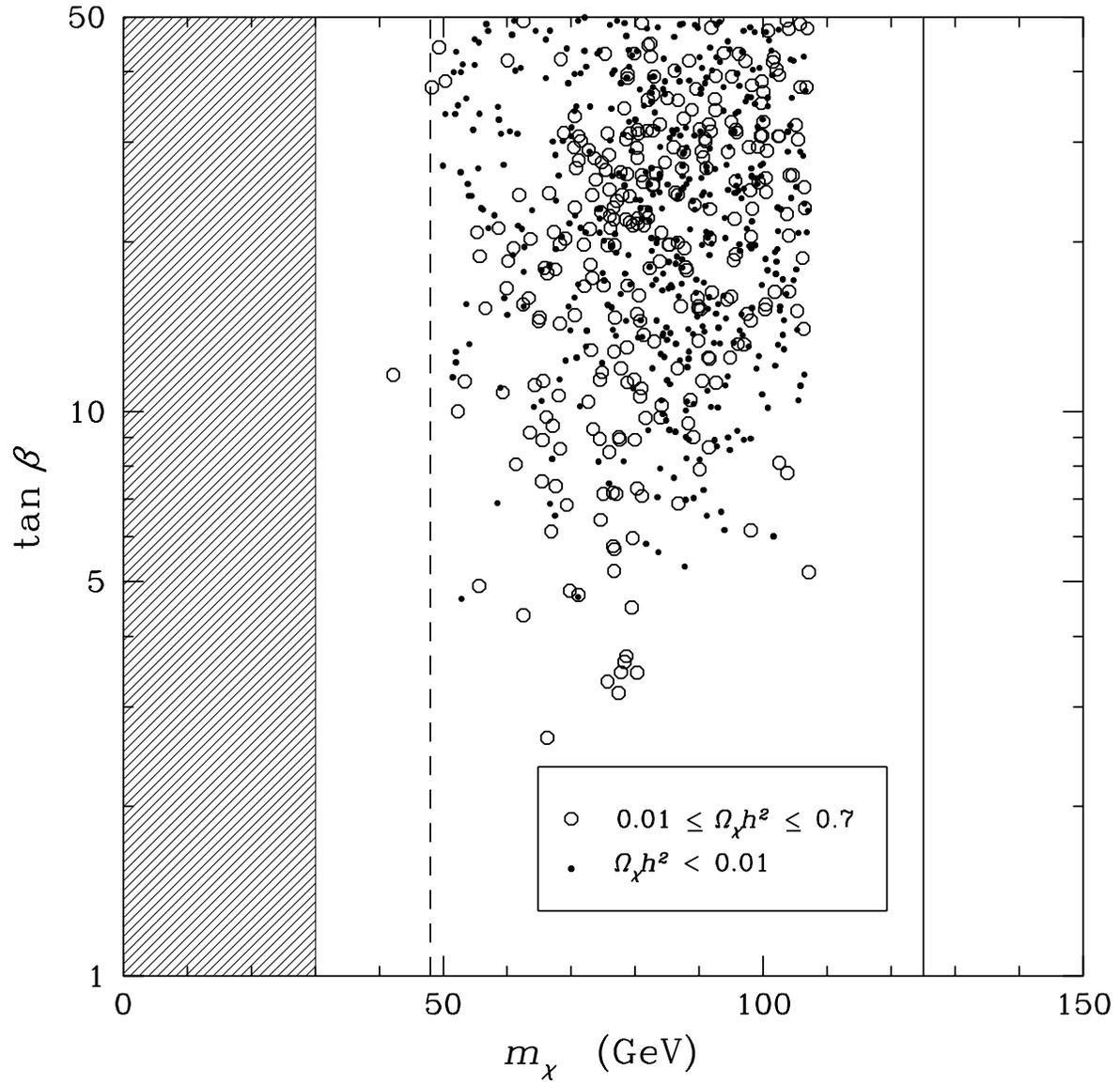
{\bf Figure 7} -- Scatter plot for set $S$ in the plane 
$m_{\chi}$ -- $\tan \beta$. 
The hatched region on the left is 
excluded by present LEP data. The dashed and the 
solid vertical lines denote the reach of LEP2 and TeV33, respectively. 

{\bf Figure A.1} -- Regions of the parameter space defined in Sect. II, 
which satisfy all accelerator constraints (including 
$b \rightarrow s + \gamma$) and the further requirement that 
$m_h$ is below some arbitrarily fixed value $m_h^*$. 
The various lines denote the following representative values of 
$m_h^*$: $m_h^* = 80, 90, 100, 110$ GeV. 
The allowed regions are given by 
the domains on the left of the various curves for each value
of $m_h^*$.

{\bf Figure A.2} -- Allowed region in the plane 
 $m_{\tilde t_1}$ -- $\tan \beta$ when 
the plot of Fig. A.1 is combined with the one of Fig. 5(a). 
The hatched region on the left is excluded by LEP data 
(without any restriction on other masses). 

\vfill\eject


\begin{thebibliography}{99}

\bibitem{drukier} A.K. Drukier, K. Freese and D.N. Spergel, Phys. Rev. 
{\bf D33}, 3495 (1986).

\bibitem{freese} K. Freese, J. Frieman and A. Gould, Phys. 
Rev. {\bf D37}, 3388 (1988).

\bibitem{dama2}      R. Bernabei et al., Roma II University preprints: \\
                     ROMA2F/98/27 and ROMA2F/98/34, August 1998.

\bibitem{dama1}      R. Bernabei et al.,
                     Phys. Lett. {\bf B424}, 195 (1998).

\bibitem{damapsa}    R. Bernabei et al., Phys. Lett. {\bf B389}, 757 (1996).

\bibitem{limiti}     A. Bottino, F. Donato, G. Mignola, S. Scopel, P. Belli 
                     and A. Incicchitti, 
                     Phys. Lett. {\bf B402}, 113 (1997).

\bibitem{our1}   A. Bottino, F. Donato, N. Fornengo and S. Scopel, Phys. 
                  Lett. {\bf B423}, 109 (1998).

\bibitem{our2}   A. Bottino, F. Donato, N. Fornengo and S. Scopel,  
                 Torino University preprint DFTT 61/97, 1997,
                 {\tt hep-ph/9710295}. 
                 

\bibitem{cm} For updated reviews on direct search experiments see, for
                instance, D.O. Caldwell and A. Morales, {\it Proceedings of
                TAUP97}, Nucl. Phys. 
                {\bf B70} (Proc. Suppl.), 43, 54 (1999).

\bibitem{susy}      H.P. Nilles, Phys. Rep. {\bf 110}, 1 (1984);
                    H.E. Haber and G.L. Kane, Phys. Rep. {\bf 117}, 75 (1985);
                    R. Barbieri, Riv. Nuovo Cim. {\bf 11}, 1 (1988).

\bibitem{carena} M. Carena, M. Quir\'os and C.E.M. Wagner, 
                    Nucl. Phys. {\bf B461}, 407 (1996).

\bibitem{haber} H.E. Haber, Z. Phys. {\bf C75}, 539 (1997).

\bibitem{bbefms1}  V. Berezinsky, A. Bottino, J. Ellis, N. Fornengo, G. Mignola
                    and S. Scopel, Astropart. Phys. {\bf 5}, 1 (1996).

\bibitem{lep2}  M. Felcini,  Workshop ``From Planck Scale to
                Electroweak Scale" , Kazimierz, 1998; 
                A. Lipniacka, {\it ibidem};
                ALEPH Collaboration, CERN--EP/98--077, 1998; 
                L3 Susy Group, L3 Note 2238. 
 
\bibitem{ichep}     K. Desch  and P. Rebecchi, talks at
                    International Conference on High Energy Physics, 
                    Vancouver, 1998.

\bibitem{bertolini} S. Bertolini, F. Borzumati, A. Masiero and G. Ridolfi, 
                    Nucl. Phys. {\bf B353}, 591 (1991).

\bibitem{bg} R. Barbieri and G.F. Giudice, Phys. Lett. {\bf B309}, 86 (1993).

\bibitem{garisto} R. Garisto and J.N. Ng, Phys. Lett. {\bf B315}, 372 (1993).

\bibitem{borzumati} F.M. Borzumati, M. Drees and M.M. Nojiri, 
                    Phys. Rev. {\bf D51}, 341 (1995).

\bibitem{wu} J. Wu, R. Arnowitt and P. Nath, Phys. Rev. {\bf D51}, 1371 (1995).

\bibitem{barger} V. Barger, M.S. Berger, P. Ohmann and R.J.N. Phillips, Phys.
                Rev. {\bf D51}, 2438 (1995).

\bibitem{chetyrkin} K. Chetyrkin, M. Misiak and M. M\"unz, Phys. Lett. 
                    {\bf B400}, 206 (1997).

\bibitem{ciuchini1} M. Ciuchini, G. Degrassi, P. Gambino and G.F. Giudice,
                    CERN preprint CERN--TH/97--279, 1998, 
                    {\tt hep-ph/9710335}. 

\bibitem{cza} A. Czarnecki and W.J. Marciano, 
              Brookhaven National Laboratory 
              preprint BNL--HET--98/11, 1998, {\tt hep-ph/9804252}.

\bibitem{ciuchini2} M. Ciuchini, G. Degrassi, P. Gambino and G.F. Giudice,
                    CERN preprint CERN--TH/98--177, 1998, {\tt hep-ph/9806308}.


\bibitem{glenn} S. Glenn (CLEO Collaboration), preprint CLEO CONF 98--17, 1998;
                 International Conference on High Energy Physics, 
                    Vancouver, 1998, paper 1011.

\bibitem{barate} R. Barate et al. (ALEPH Collaboration), CERN preprint
                CERN--EP/98--044, 1998.

\bibitem{age} B. Chaboyer, P. Demarque, P. Kernan and L.M. Krauss, 
              Astrophys. J. {\bf 494}, 96 (1998).

\bibitem{hubble} A. Sandage et al., Astrophys. J. Lett. {\bf 460}, L15 (1996); 
            W. L. Freedman,  talk at the 18th Texas Symposium, 
            Chicago, 1996,
            {\tt astro-ph/9706072} and references quoted therein.

\bibitem{ouromega}   A. Bottino, V. de Alfaro, N. Fornengo, G. Mignola 
                     and M. Pignone,
                     Astropart. Phys. {\bf 2}, 67 (1994).


\bibitem{coannih}  P. Bin\'{e}truy, G. Girardi and P. Salati, Nucl. Phys. 
                    {\bf B237}, 285 (1984);
                    K. Griest and D. Seckel,  Phys. Rev. {\bf D43}, 3191 (1991);
                    S. Mizuta and M. Yamaguchi, 
                     Phys. Lett. {\bf B298}, 120 (1993); 
                     J. Edsj\"o and P. Gondolo, Phys. Rev. {\bf D56}, 
                     1879 (1997).

\bibitem{turner1}   E. Gates, G. Gyuk and M.S. Turner,
                     Phys. Rev. Lett. {\bf 74}, 3724 (1995);
                     Astrophys. J. Lett. {\bf 449}, L123 (1995);
                     Phys. Rev. {\bf D53}, 4138 (1996).

\bibitem{turner}    E. Gates, G. Gyuk and M.S. Turner,
                     talk presented at 18th Texas Symposium on 
                     Relativistic Astrophysics, 
                     Chicago, 1996, {\tt astro-ph/9704253}.


\bibitem{barbieri} R. Barbieri, M. Frigeni and G.F. Giudice,
                   Nucl. Phys. {\bf B313}, 725 (1989).

\bibitem{griest}   K. Griest, Phys. Rev. {\bf D38}, 2357 (1988); 
                   Phys. Rev. Lett. {\bf 61}, 666 (1988).
 
\bibitem{svz}     M.A. Shifman, A.I. Vainshstein and V.I. Zacharov, 
                    Phys. Lett. {\bf B78}, 443 (1978);
                    JEPT Lett. {\bf 22}, 55 (1975).

\bibitem{gasser} J. Gasser, H. Leutwyler and M. E. Sainio,
                 Phys. Lett. {\bf B253}, 252 (1991).


\bibitem{liu} S. J. Dong and K. F. Liu in {\it Proceedings of Lattice 94}, 
              Nucl. Phys. {\bf B42} (Proc. Suppl.), 322 (1995).

\bibitem{bj} J. Bijnens, J. Prades and E. de Rafael, 
                Phys. Lett. {\bf B348}, 226 (1995).

\bibitem{gaisser} T.K. Gaisser, G. Steigman and S. Tilav, 
                  Phys. Rev. {\bf D34}, 2206 (1986).

\bibitem{bbefms} V. Berezinsky, A. Bottino, J. Ellis, N. Fornengo, G. Mignola
                  and S. Scopel, Astropart. Phys. {\bf 5}, 333 (1996).

\bibitem{omegamatter} W.L. Freedman, R. Kirshner and C. Lineweaver,
                    talks given at the International Conference of 
                    Cosmology and Particle Physics, Geneva, 1998,
                    http://wwwth.cern.ch/capp98/programme.html; 
                    M. White, {\tt astro-ph/9802295}; 
                    N.A. Bahcall and X. Fan, {\tt astro-ph/9804082};
                    C. Lineweaver, {\tt astro-ph/9805326}.

\bibitem{tev} Report of the TeV--2000 Study Group, D. Amidei, R. Brock (Eds.), 
            Fermilab preprint FERMILAB--PUB--96/082, 1996.

\bibitem{baer} H. Baer, B.W. Harris and X. Tata, Florida State University
                        preprint FSU--HEP--980626, 1998, 
                        {\tt hep-ph/9807262}.

\bibitem{bsg} The relevance of the $b \rightarrow s + \gamma$ constraint 
 for WIMP direct detection rates (without considering annual 
modulation effect) has been discussed in the past in a number of
 papers with some conflicting conclusions. See for example: 
                F.M. Borzumati, M. Drees and M.M. Nojiri, 
                Phys. Rev. {\bf D51}, 341 (1995);
                P. Nath and R. Arnowitt, Phys. Rev. Lett. {\bf 74}, 4592 (1995);
                V. Berezinsky, A. Bottino, J. Ellis, N. Fornengo, G. Mignola
                and S. Scopel, Astropart. Phys. {\bf 5}, 1 (1996);
                L. Bergstr\"{o}m and P. Gondolo,
                Astropart.  Phys. {\bf 5}, 263 (1996); 
                V.A. Bednyakov, S.G. Kovalenko, H.V.
                Klapdor--Kleigrothaus and Y. Ramachers, 
                Z. Phys. {\bf A357}, 339 (1997).

\bibitem{pierre} A. Bottino, F. Donato, N. Fornengo and P. Salati, 
                 Phys. Rev. D (to be published)
                and references quoted therein, {\tt astro-ph/9804137}. 

\bibitem{bess} H. Matsunaga et al. (BESS Collaboration), in  Proceedings 
               of the 25th International Conference of Cosmic Rays,
               Durban, 1997.

\bibitem{ams} Ahlen et al., Nucl. Instr. Methods {\bf A350}, 351 (1994).

\bibitem{macro} M. Ambrosio et al. (MACRO Collaboration), MACRO/PUB1/98; 
                T.  Montaruli et al., {\it Proceedings of TAUP97}, Nucl. Phys. 
                {\bf B70} (Proc. Suppl.), 367 (1999).


\bibitem{baksan} M.M. Boliev et al., {\it Proceedings of TAUP97}, Nucl. Phys.
                {\bf B70} (Proc. Suppl.), 371 (1999).

\bibitem{indirectnew} A. Bottino, F. Donato, N. Fornengo and S. Scopel, 
            Torino University preprint DFTT 49/98, 1998 (to appear).

\bibitem{companion} A. Bottino, F. Donato, N. Fornengo and S. Scopel, 
            Torino University preprint DFTT 48/98, 1998.

\end{thebibliography}
\end{document}